\documentclass[prd,twocolumn,showpacs]{revtex4}

\usepackage{amsmath}
\usepackage{amssymb}
\usepackage{bm}
\usepackage{epsf}
\usepackage[all, knot]{xy}

\xyoption{arc}
\begin{document}

\title{Black hole entropy in modified gravity models}

\author{F.~Briscese$\,^{a}$\footnote{E-mail: briscese@dmmm.uniroma1.it} and
E.~Elizalde$\,^{b}$\footnote{E-mail: elizalde@ieec.uab.es}}

\affiliation{$^{a}$~Dipartimento di Modelli e Metodi Matematici and GNFM \\
Universit\`a degli Studi di Roma ``La Sapienza", Via A. Scarpa 16,
I-00161, Roma, Italy\\
$^{b}$~Institut de Ci\`encies de l'Espai (CSIC) \& IEEC-CSIC,
Campus UAB \\ Facultat de Ci\`encies, Torre C5-Parell-2a pl,
E-08193 Bellaterra (Barcelona) Spain}

\begin{abstract}
\begin{center}
{\bf Abstract}
\end{center}
An analysis of some modified gravity models, based on the study of
pure Schwarzschild and of Schwarzschild-de Sitter black holes, and
involving the use of the Noether charge method, is carried out. Corrections to the classical Einsteinian black hole entropy appear. It is shown explicitly how the condition of positive entropy can be used in order to constrain the viability of modified gravity theories.
\end{abstract}


\maketitle

\section{Introduction.}
Increasing interest is attracted by modified versions
of general relativity \cite{od1}. They have been proposed as
 serious alternatives to Einstein's theory of gravitation, and could
be used to describe more accurately the observed accelerated expansion
of our universe \cite{od2,all,Capo,Fay,od5}. In addition,
it has been shown  \cite{od8} that it is actually possible to reconstruct the explicit
form of the postulated curvature function $f(R)$, from the universe expansion history.

It is quite well known that modified versions of general relativity
are mathematically equivalent to scalar fields models (see e.g.
\cite{od1}), meaning that a solution in a modified gravity model can
always be mapped into a solution of the corresponding scalar field
theory. In spite of this mathematical correspondence, physical
equivalence does not always follow. In fact, two corresponding
solutions of two equivalent theories can actually exhibit rather
different physical behaviors. Furthermore, it is not necessary, in
order to justify modified gravity, to do it by always using this relation
with scalar field theories. Because of the new situation, in the
following we will disregard this mathematical equivalence and do
consider in our analysis modified gravity as an independent theory
aiming directly at some measurable physical properties. What is
more, in our treatment modified gravity will in fact be viewed just as a
different classical theory of gravitation.

Although other models have been considered \cite{od3} with the
Gauss-Bonnet scalar in the action, here we shall restrict our
attention to pure $f(R)$ models. As often discussed, there are
limitations on the function $f(R)$ when trying to construct a theory
which is in agreement with the very precise solar system tests
carried out so far, as well as with all the known cosmological
bounds \cite{all}. Recently, different models of that kind have been
studied \cite{od8,od2,Hu,ST,AB}, the last three of them having been
reported to pass all solar system tests; in addition, they exhibit a
number of very interesting features. In \cite{od10} possible Newton
law corrections to such models have been considered.

We present here an analysis of the models above based on the study
of pure Schwarzschild and also Schwarzschild-de Sitter black holes
(SBH, SdSBH), calculated through use of the Noether charge method.
We start with a discussion of two examples considered in
\cite{od5,od6} and go over to study more recent ones
\cite{Hu,ST,AB}. The direct confrontation of basic quantities, as
the black hole entropy, with the well established, classical
(Einsteinian) result can offer a further insight into the
construction of a general $f(R)$ theory. We start with a short
review of modified $f(R)$ gravity and the Noether charge method to
compute the BH entropy and then calculate the BH entropy for the
models in Refs.~\cite{od8,od2}. After that, we extend our analysis
to the models \cite{Hu,ST,AB}, taking due care of the sign of the BH
entropy and also discussing the stability conditions as well as the
existence of a Schwarzschild BH solution. We conclude by providing a
brief comparison of the different models considered.
\medskip

\section{Black hole entropy in modified gravity.}
The action for $f(R)$ gravity (see e.g. \cite{od1} for a review) is
\begin{equation}
J = \frac{1}{k^2}\int d^4x \sqrt{-g}\mathcal{L} =
\frac{1}{k^2}\int d^4x \sqrt{-g}\left( R + f(R) \right),
\end{equation}
with $f(R)$ a generic function. As discussed in \cite{od2}, in order
to give rise to a realistic cosmology this function needs to fulfill
some limiting conditions. However, for the moment we can ignore
them, because they do not affect the considerations which will
follow. The equations of motion for this theory, in the presence of
matter, are
\begin{equation}
\begin{array}{rl}
\left(1+ f'(R) \right) R_{\mu \nu} - \frac{1}{2}\left(R + f(R)
\right) g_{\mu \nu} + g_{\mu \nu} \Box f'(R) \\ -
\bigtriangledown_\mu \bigtriangledown_\nu  f'(R)= k^2 T_{\mu
\nu},
\end{array} \label{eq2}
\end{equation}
where $T_{\mu \nu}$ is the matter stress energy tensor. Contracting
the indices in the last equation, we obtain the relation
\begin{equation}
R \left( f'(R) - 1 \right) -2 f(R) + 3 \Box f'(R) = k^2 T^\mu _\mu.
\end{equation}
Since we are interested in the study of the Schwarzschild solution
(in the case when there is a $R=0$ solution) or either in
Schwarzschild-De Sitter black holes, we  restrict our reasoning to
the case of metric tensors with constant scalar curvature in the
vacuum. In that case, we simply have \cite{foot1}
\begin{equation}
R_0 \left( f'(R_0) - 1 \right) -2 f(R_0) = 0.
\end{equation}
For completeness, let us recall that, in order to build a realistic
modified gravity, $f(R)$ needs satisfy the two conditions:
\begin{equation} \lim_{R \rightarrow
\infty}f(R) = \mbox{const},
\qquad \lim_{R \rightarrow 0}f(R) = 0,
\end{equation}
The first condition corresponds to the existence of an effective cosmological constant at high curvature. The second one allows for vacuum solutions, as for example Minkowski or Schwarzschild space-times. Then, although an effective cosmological constant exists, vacuum solution are preserved and it is legitimate to study them also in a large scale universe with nonzero $R$. Moreover, in order to
give rise to stable solutions \cite{Saw}, the following additional condition needs to be fulfilled:
\begin{equation} [1+f'(R)]/f''(R) > R.
\end{equation}
We can now
consider the Schwarzschild-De Sitter metric, a spherically symmetric solution of (\ref{eq2}) with constant curvature $R_0$ (see
\cite{Multa})
\begin{equation}
ds^2 = a(r) dt^2 -  dr^2/a(r) - r^2 d\Omega,
\end{equation}
where $ a(r) = 1 -2m/r - R_0 r^2/12$ and $R_0 \left( f'(R_0) - 1 \right) -2 f(R_0) = 0 $.

We will use the Noether charge method, as discussed in
\cite{od3,od4}, in order to calculate the entropy for the
Schwarzschild-De Sitter BH. The entropy formula reads
\begin{equation}
S = 4 \pi \int_{S^2} \sqrt{-g}\ \frac{\partial
\mathcal{L}}{\partial R},
\end{equation}
and the integration is made on the external horizon of events
surface. In the case of constant curvature, for a generic modified
theory, the result is
\begin{equation}
S = [1+ f'(R_0)] A_H/4G,
\end{equation}
where $A_H$ is the area of the BH horizon. This enables
us to calculate the BH entropy for a generic $f(R)$ theory.
We must here stress the fact that the requirement of positive black hole entropy simply avoids the appearance of ghost or tachyon fields in the corresponding scalar field theory. Then a negative entropy is simply a footprint of some instabilities in the Einstein frame. What is new in this picture is that we do not need to involve the (mathematical) equivalence of these  models in order to give a physically meaningful interpretation of such constrain.

\medskip

\section{Black hole entropy for two modified gravity models
with no $R=0$ solutions.} In order to illustrate the method with
explicit examples of entropy calculation, we analyze here two
modified gravity models that appeared some time ago and which have
been quite successful up to now. In those models no $R=0$ solution occurs. The
first one, introduced in \cite{od5}, is given by
\begin{equation}
f(R) = - a \left(R-\Lambda_1\right)^{-n} + b \left( R -
\Lambda_2 \right)^m,
\end{equation}
with $ m,n,a,b >0$. The condition to obtain a SdSBH  (namely,
$2f(R_0)= R_0 \left(f'(R_0)-1 \right)$) leads to
\begin{eqnarray}
&& R_0 \left[ an\left(R_0- \Lambda_1 \right)^{-n-1} + b m
\left( R_0 - \Lambda_2 \right)^{m-1} -1 \right] \nonumber \\ && \hspace*{10mm} = 2 \left[ -
a\left(R_0 - \Lambda_1 \right)^{-n} + b \left( R_0 -
\Lambda_2 \right)^m \right],
\end{eqnarray}
so that we have for the entropy
\begin{equation}
S = \frac{A_H}{4G} \left[ 1+ n a\left(
R_0-\Lambda_1\right)^{-n-1} + m b \left(R_0 -\Lambda_2
\right)^{m-1} \right].
\end{equation}
Thus the  SdSBH entropy  is positive for all $R_0 >
\Lambda_1,\Lambda_2$.

The second model, studied in \cite{od6}, is defined by
\begin{equation}
f(R) = \alpha \ln(R/\mu^2) + \beta R^m.
\end{equation}
This modified gravity model does not admit  vacuum solutions, thus
we can calculate the entropy for the SdSBH. The Ricci scalar is such
that it satisfies the relation
\begin{equation}
2\alpha \ln (R_0/\mu^2) + \beta (2-m) R_0^m - \alpha =0.
\end{equation}
The SdSBH entropy is given by
\begin{equation}
S = \frac{A_H}{4G} \left( 1 + \alpha/R_0 + \beta m R_0^{m-1}
\right),
\end{equation}
and turns out to be positive for all values of $R_0>0$.
\medskip

\section{Black hole entropy in modified gravity models
that comply with the solar system tests.} We now analyze three
recent models \cite{Hu,ST,AB} which have been proven to comply with
the solar-system as well as with other cosmological parameter
constraints. Their respective authors have given a complete
discussion of each model, taking care to provide a range for the
free parameters contained in the $f(R)$ function, and have also
produced stable solutions. Here we just want to stress, with the
help of these examples, how the corresponding BH entropy calculation
offers a further tool in order to confront each of those modified
gravity theories with Einstein's general relativity, given the fact
that the presence of spherically symmetric BH solutions is a
necessary element of all local tests.
\medskip

\subsection{The Hu-Sawicki model.}
In this model \cite{Hu} (with $n,c_1,c_2>0$) we have
\begin{equation}
f(R) = - m^2 c_1 \frac{\left( \frac{R}{m^2} \right)^n}{c_2
\left(\frac{R}{m^2} \right)^n +1}.
\end{equation}
In \cite{Hu}, $m^2$ is chosen such that, at cosmological scale, $R>>m^2$ at the present epoch, and $f(R)$ satisfies the condition $f''(R)>0$ for $R>>m^2$. This also ensure that solutions with $R>>m^2$ are stable. Moreover the requirement that $ c_1/c^2_2 \rightarrow 0 $ at fixed $c_1/c_2$ gives a cosmological constant, in both cosmological and local tests of gravity. In spite of this fact, since $f(0)=0$, this theory admits the Schwarzschild solution (i.e. $R=0$). By the way we note that the stability condition for the vacuum solution is not satisfied unless $n=1$ and $1-c_1>0$. Therefore, except of this case, vacuum solutions (than also SBH) are unstable. Note also that $n=1$ corresponds to a Lagrangian $ {\it L}= R (1-c_1)/k^2 $ for small $R$, so it is associated with a correction to the gravitational coupling constant for small $R$, giving an effective $G_{eff.} = G/(1+f'(0))$ (see \cite{Vollick}).

The entropy formula gives for the SdS metric
\begin{equation}
\label{entropyhu}
S(R_0) = \frac{A_H}{4G}
\left(1- n c_1  \frac{\left( \frac{R_0}{m^2} \right)^{n-1}}{\left(c_2  \left(\frac{R_0}{m^2} \right)^n +1\right)^2}
\right).
\end{equation}
The entropy for the Schwarzschild solution is
\begin{equation}
\begin{array}{ll}
S(0) = \left(1-c_1 \right)\frac{A}{4G}$,  for  $n=1;\\
S(0) = \frac{A}{4G}  $,   for    $ n>1 $,      as in the Einstein theory;        $\\
S(0) =  - \infty  $,   for    $  0<n<1 ,  c_1>0.
\end{array}
\end{equation}
Then, in the only stable case, with $n=1$ and $1-c_1>0$, a correction to the classical Einstenian BH entropy is found. From (\ref{entropyhu}) it also follows that, for SdS BH with $R_0>>m^2$, the entropy is positive and corrections to its Einstenian value are of order $(m^2/R_0)^{n+1}$.
\medskip

\subsection{The Starobinsky model.}
In this model $f(R)$ is \cite{ST}
\begin{equation}
f(R) = \lambda C \left\{ \left[ 1+(R/C)^2
\right]^{-n} -1 \right\},
\end{equation}
from where \begin{equation} f'(R) = -2 n \lambda (R/C) \left[
1+(R/C)^2 \right]^{-n-1}. \end{equation} Note that in this case
$f''(0) < 0$ an thus, although this model admits a SBH solution, it
is unstable together with all its vacuum solutions. In \cite{ST} the author limits his analysis to solutions that satisfy the following stability conditions
\begin{equation}
\label{stabilitistarobinski}
1+f'(R)>0 \, , \, f''(R)>0 \, , \, 1+f'(R)> R f''(R)
\end{equation}
We can therefore
consider the SdSBH solutions, with curvature $R_0$ given by
\begin{eqnarray}
&& R_0 \left\{ 2 n \lambda (R_0/C) \left[
1+(R_0/C)^2 \right]^{-n-1} +1 \right\}\nonumber \\ && \hspace*{10mm}  = - 2
\lambda C \left[ \left( 1+(R_0/C)^2
\right)^{-n} -1 \right],
\end{eqnarray}
that satisfies also (\ref{stabilitistarobinski}).
In this case, the entropy is just
\begin{equation}
S = \frac{A_H}{4G} \left\{1-2 n \lambda (R_0/C)
\left[ 1+(R_0/C)^2 \right]^{-n-1}     \right\}.
\end{equation}
Therefore, in this case a {\it non trivial} correction of the SdSBH entropy
is found. We just stress the fact that the SBH
solutions have classical entropy but are {\it unstable}, and that the SdSBH ones
have a modified entropy which, under the limitations stated in \cite{ST}, is strictly positive.
\medskip

\subsection{The Appleby-Battye model.}
Here $f(R)$ is given by \cite{AB}
\begin{equation}
f(R) = -R/2 + \log \left[ \cosh(aR) -
\tanh(b) \sinh(aR)\right]/2a
\end{equation}
and
\begin{equation}
f'(R) = \left[- 1+\tanh(aR-b)\right]/2. \end{equation} This model
admits a SBH solution. The entropy for the SdSBH is simply
\begin{equation} S = \left[ 1+ \tanh(aR_0-b) \right]
A_H/8G.
\end{equation}
We can use the stability condition given in \cite{AB},
$aR_0-b>> 1$, to obtain
\begin{equation}
S \simeq  A_H/4G.
\end{equation}
For  the SBH, the stability condition is just $b<<0$.
Moreover, being $f''(R)>0$ for all $R$, in this model the vacuum
solutions are always stable and there are no substantial corrections
to the classical result.
\medskip

\subsection{Comparison of the behavior of
$f(R)$ for the different models.} It is interesting to put together
all three models and explicitly compare the behavior of the function
$f(R)$, in particular, the stability of the Euclidean limit and the
asymptotic behavior at large curvature.   To simplify the
comparison, we do not play with the values of the different
parameters and set all coefficients equal to 1 and the curvature
powers equal to 2 or 4. For the case of the Hu-Sawicki model, with
\begin{equation}f_{HS} (x) = -\frac{x^4}{1+x^4}, \end{equation} being $x\equiv
R/m^2$, the corresponding plot is given in Fig.~1.
\begin{figure}
\centerline{\epsfxsize=8cm \epsfbox{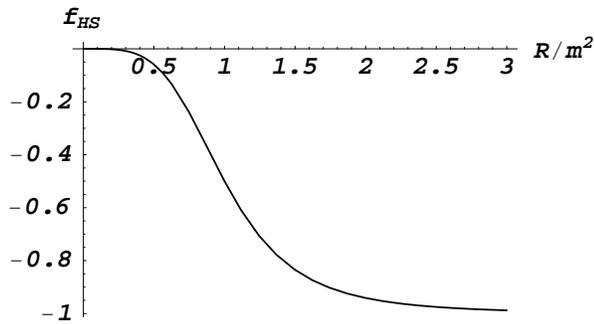}}
\caption{{\protect\small  Plot of the function f(R) for
the Hu-Sawicki model.}} \label{f1}
\end{figure}
For the case of the Starobinsky model, with \begin{equation} f_{HS}
(x) = -1 + \frac{1}{(1+x^2)^2}, \end{equation} being $x\equiv
R/R_0$, we obtain Fig.~2.
\begin{figure}
\centerline{\epsfxsize=8cm \epsfbox{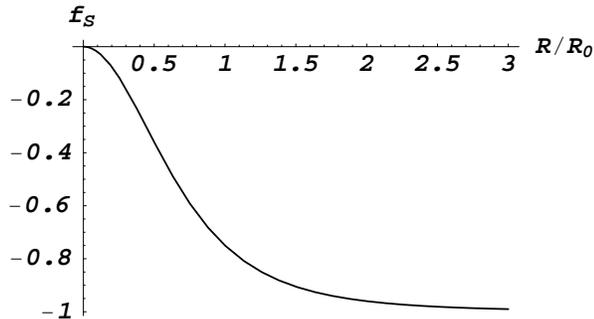}}
\caption{{\protect\small  Plot of the function f(R) for
the  Starobinsky model.}} \label{f2}
\end{figure}
And for the case of the Appleby-Battye model, with
\begin{equation} f_{AB} (x) = -\frac{x}{2} + \frac{1}{2} \log ( \cosh
x + \sinh x), \end{equation}  being $x\equiv aR$ and $b=1/2$, Fig.~3.
\begin{figure}
\centerline{\epsfxsize=8cm \epsfbox{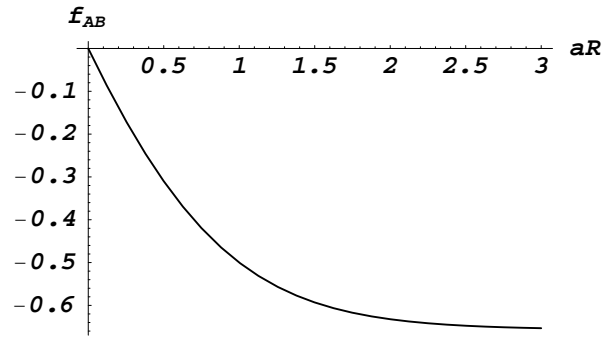}}
\caption{{\protect\small  Plot of the function f(R) for
the Appleby-Battye  model.}} \label{f3}
\end{figure}

In a first comparison of these different models, we note that the one of
Starobinsky, as remarked by the author himself \cite{ST}, has
unstable vacuum solutions. This seems  true also for the Hu-Sawicki
model, except for the case when $0<n<1$, that leads to a negative and
infinite BH entropy. The Appleby-Battye model has the important
property to possess stable vacuum solutions for a suitable range of
the free parameters. It also yields an unmodified expression of the
BH entropy.
\medskip

\section{BH entropy in a new model that unifies inflation and cosmic
acceleration.} Very recently, a modified gravity model has been
published \cite{od0707}, that unifies inflation and cosmic
acceleration under the same picture and also complies with the solar
system tests. It is
\begin{equation}
f(R) = - f_0 \int_0^R \exp\left[-\alpha
\frac{R_1^{2n}}{\left(x-R_1\right)^{2n}}-f_0 \frac{x}{\Lambda_i}\right]
dx,
\end{equation}
where $0<f<1$ and $R_1$ is a constant given by $ f_0 R_1$ $ \int_0^1
e^{-\alpha/x^2} dx  = R_{now},$ and $R_{now}$ is the Ricci scalar at
present. The effective cosmological constant in the early universe
is simply $-f(-\infty) = \Lambda_i$ and the present cosmological
constant is $2R_0$. Because of the fact that $f(0)=0$, this model
allows for SBH. To be general, we first calculate the SdSBH entropy,
which is given by
\begin{equation}
S = \frac{A_H}{4G}  \left\{ 1 - f_0 \exp\left[-\alpha
\frac{R_1^{2n}}{\left(R_0-R_1\right)^{2n}}-f_0
\frac{R_0}{\Lambda_i}\right] \right\},
\end{equation}
where $R_0$ is the SdSBH curvature that fulfills condition (4).
Thus, black holes are less entropic than in Einstein's theory. For
the SBH, we have
\begin{equation}
S = \frac{A_H}{4G}  \left( 1 - f_0 e^{-\alpha } \right).
\end{equation}
Note that the stability condition  is here
$f''(R)>0$, thus
\begin{equation}
\frac{f_0}{\Lambda_i} > 2n \frac{R_1^{2n}}{\left(R_0 - R_1
\right)^{2n+1}}.
\end{equation}
To have stability for SBH,  we need that
\begin{equation}
n < \frac{R_1}{2} \frac{f_0}{\Lambda_i}.
\end{equation}
This can be considered, together with the condition $n>10-12$ stated
in \cite{od0707}, to avoid Newton law corrections in our solar
system and on the earth surroundings.
\medskip

\section{Conclusions.}
Comparison of BH entropy in modified gravity theories and in the usual Einstenian gravity case have been carried out.
We have  here analyzed different suitable models, recently considered in the literature, and have shown explicitly how  corrections to the `classical' BH entropy can in fact appear. We have also argued that the condition of positive entropy can be used as an extra condition in order to constrain the viability of modified gravity theories. Of course this conditions is equivalent to the requirement that neither ghost nor tachyon fields appear in the equivalent scalar field models. Anyhow, if referred to the BH entropy, this condition has a direct interpretation in the framework of modified gravity, without needing to pass through the (mathematically but not physically equivalent) scalar field theories. We hope that this quite simple considerations may be useful for future analysis.
\medskip

We thank Sergei Odintsov for helpful discussions. This investigation is 
partly based on work done by EE while on leave at the Department of Physics and
Astronomy, Dartmouth College, 6127 Wilder Laboratory, Hanover, NH
03755, USA, and was completed during a stay of FB in Barcelona. 
It has been supported by MEC (Spain), projects FIS2006-02842,
and by AGAUR 2007BE-1003 and contract 2005SGR-00790.


\addcontentsline{toc}{chapter}{Bibliography}

\begin{thebibliography}{9}

\bibitem{od1} S. Nojiri and S.D. Odintsov,
Int. J. Geom. Meth. Mod. Phys. {\bf 4}, 115 (2007).
\bibitem{od2} S. Nojiri and S.D. Odintsov.  Phys. Rev. {\bf D74},  086005                                                                                                      (2006); S. Capozziello, S. Nojiri, S.D. Odintsov and
A. Troisi, Phys. Lett. {\bf B639}, 135 (2006).

\bibitem{all}
F. Faraoni, arXiv:gr-qc/0607116; arXiv:gr-qc/0511094;
A. Cruz-Dombriz and A. Dobado,
arXiv:gr-qc/0607118; N. Poplawski,
gr-qc/0610133; A. Brookfield, C. van de Bruck and L. Hall,
arXiv:hep-th/0608015;  T. Sotiriou and S. Liberati,
arXiv:gr-qc/0604006;  A. Bustelo and D. Barraco,
arXiv:gr-qc/0611149; G. Olmo, arXiv:gr-qc/0612047;
F. Briscese, E. Elizalde, S. Nojiri
and S.D. Odintsov, Phys. Lett. {\bf B646}, 105 (2007);
B. Li and J. Barrow, arXiv:gr-qc/0701111;  O. Bertolami,
C. Boehmer, T. Harko and F. Lobo, arXiv:0704.1733;
S. Carloni, A. Troisi and P. Dunsby,
arXiv:0706.0452. S. Nojiri and S. Odintsov,
Phys. Lett. {\bf B599}, 137 (2004);  M. Abdalla, S. Nojiri
and S.D. Odintsov, Class. Quant. Grav. {\bf 22}, L35 (2005); G. Cognola, E. Elizalde, S. Nojiri, S.
D. Odintsov and S. Zerbini, JCAP {\bf 0502}, 010 (2005); Phys. Rev. {\bf D73},
084007 (2006); D.A. Easson, Int. J. Mod. Phys. {\bf A19},
5343 (2004); S. Capozziello, V. Cardone
and A. Troisi, Phys. Rev. {\bf D71}, 043503 (2005) ; G. Allemandi, A. Borowiec,
M. Francaviglia and S.D. Odintsov, Phys. Rev. {\bf D72},  063505 (2005);   J. Kainulainen, M. Juvela
and J. Alves, arXiv:astro-ph/0703607;
J.A.R. Cembranos, Phys. Rev. {\bf D73}, 064029 (2006);
T. Koivisto and H. Kurki-Suonio,
arXiv:astro-ph/0509422; T. Clifton and J. Barrow,
arXiv:gr-qc/0509059; O. Mena, J. Santiago and J. Weller,
arXiv:astro-ph/0510453;  I. Brevik, arXiv:gr-qc/0601100;
R.Woodard, arXiv:astro-ph/0601672; T. Koivisto,
arXiv:astro-ph/0602031; G. Cognola, M. Castaldi and S. Zerbini,
arXiv:gr-qc/0701138; S. Nojiri, S.D. Odintsov and P. Tretyakov,
arXiv:0704.2520[hep-th];  L. Amendola and S. Tsujikawa,
arXiv:0705.0396[astro-ph]; K. Uddin, J. Lidsey and R. Tavakol,
arXiv:0705.0232[gr-qc].

\bibitem{Capo} S. Capozziello, Int. J. Mod. Phys. {\bf D11}, 483 (2002); S.
Capozziello, S. Carloni and A. Troisi, Rec. Res. Develop. Astron.
Astroph.-RSP/AA/21-2003; S.M. Carroll,
V. Duvvuri, M. Trodden and S. Turner, Phys. Rev. {\bf D70}, 043528
(2004).

\bibitem{Fay} S. Fay, S. Nesseris and L.
Perivolaropoulos, arXiv:gr-qc/0703006; S. Fay, R. Tavakol and S.
Tsujikawa, arXiv:astro-ph/0701479.

\bibitem{od5} S. Nojiri, S.D. Odintsov.  Phys. Rev. {\bf D68}, 123512 (2003).

\bibitem{od8} S. Nojiri and S.D. Odintsov, {\it Modified gravity and
its recontruction from the universe espansion history},  ERE 2006;
 arXiv:hep-th/0611071.

\bibitem{od3} G. Cognola, E. Elizalde, S. Nojiri, S.D. Odintsov and S.
Zerbini. Phys. Rev. {\bf D73}, 084007 (2006).

\bibitem{Hu} W. Hu and I. Sawicki,
  arXiv:0705.1158 [astro-ph], submitted to Phys. Rev. D.

\bibitem{Vollick} D.N. Vollick, {\it Noether Charge and Black Hole Entropy in Modified Theories of Gravity}, arXiv:0710.1859.

\bibitem{ST} A.A. Starobinsky, {\it Disappearing cosmological
constant in f(R) gravity}, JETPL to appear;  arXiv:0706.2041 [astro-ph].
\bibitem{AB} S.A. Appleby and R.A. Battye,   arXiv:0705.3199v2 [astro-ph].

\bibitem{od10} S. Nojiri, S.D. Odintsov, arXiv:hep-th/07061378.

\bibitem{od6} S. Nojiri and S.D. Odintsov, Gen. Rel. Grav. {\bf 36}, 1765 (2004).

\bibitem{foot1} In order to be able to obtain the Schwarzschild
solution in modified gravity, the function $f(R)$ must satisfy
the condition $f(0) = 0$.

\bibitem{Saw} I. Sawicki and W. Hu, Phys. Rev. {\bf D75}, 127502 (2007); V. Faraoni, Phys.Rev.{\bf D72},061501,(2005).

\bibitem{Multa} T. Multamaki and I. Vilja, Phys. Rev. {\bf D74}, 064022 (2006);
arXiv:astro-ph/0606373; arXiv:astro-ph/0612775.

\bibitem{od4} I. Brevik, S. Nojiri, S.D. Odintsov and  L. Vanzo,
Phys. Rev. {\bf D74}, 064022 (2006).

\bibitem{od7} S. Nojiri and S.D. Odintsov, Phys. Rev. {\bf D74}, 086005 (2006).

\bibitem{od0707} S. Nojiri and S.D. Odintsov, arXiv:hep-th/0707.1941.

\end{thebibliography}
\end{document}